\documentclass[12pt,twoside]{article}
\newcommand{\be}{\begin{equation}}
\newcommand{\ee}{\end{equation}}

\newcommand{\ledd}{$L_{\rm Edd}$~}

\newcommand{\gtsima}{$\; \buildrel > \over \sim \;$}
\newcommand{\ltsima}{$\; \buildrel < \over \sim \;$}
\newcommand{\prosima}{$\; \buildrel \propto \over \sim \;$}
\newcommand{\gsim}{\lower.5ex\hbox{\gtsima}}
\newcommand{\lsim}{\lower.5ex\hbox{\ltsima}}
\newcommand{\simgt}{\lower.5ex\hbox{\gtsima}}
\newcommand{\simlt}{\lower.5ex\hbox{\ltsima}}
\newcommand{\simpr}{\lower.5ex\hbox{\prosima}}

\newcommand{\etal}{{et al.}}
\newcommand{\cxo}{\textit{Chandra~}}

\usepackage{xrb2007}
\usepackage{epsfig}
\usepackage{graphics}

\begin{document}

\session{Jets}

\shortauthor{Gallo}
\shorttitle{Jets at low and high luminosity}

\title{Multi-wavelength Observations of Jets at High and Low X-ray Luminosities}
\author{Elena Gallo}
\affil{Physics Department, University of California Santa Barbara, CA 93106 -- Chandra Fellow}

\begin{abstract}
I shall briefly review the observational properties of radio-emitting jets
from Galactic X-ray binaries in relation with other wavebands: infrared,
optical and X-ray. Special attention is paid to recent results obtained with
the Spitzer Space Telescope on quiescent black holes as well as
ultra-compact neutron star X-ray binaries.
\end{abstract}

\section{Introduction}

Traditionally, the key observational aspect of X-ray binary jets lies
in their synchrotron radio emission. Such jets appear to come in two
types: milliarcsec-scale, continuous jets with flat radio-mm spectra,
and arcsec-scale optically thin jets resolved into discrete plasmons
moving away from the binary core (Fender 2006).  These are commonly
referred to as {\it steady} and {\it transient} jets,
respectively. Over the last few years, however, observations at higher
wavelengths have added substantially to our knowledge in this
field. In the following, I shall review the properties of X-ray binary
jets in response to global changes in the accretion flow, focusing on
recent progress made in the mid-IR thanks to the unprecedented
sensitivity of the Spitzer Space Telescope.

\section{Overview: Radio Jets and X-ray States}

\subsection{Black Holes}

Radiatively inefficient hard X-ray states are associated with
flat/slightly inverted radio-to-mm spectra and persistent radio flux
levels~(Fender 2001). In analogy with compact extragalactic radio
sources~(Blandford \& K\"{o}nigl 1979), the flat spectra are thought
to be due to the superimposition of a number of peaked synchrotron
spectra generated along a conical jet, with the emitting plasma
becoming progressively thinner at lower frequencies as it travels away
from the jet base.  The jet interpretation has been confirmed by high
resolution radio maps of two hard state black hole X-ray binaries
(BHBs): Cygnus X-1~(Stirling \etal~2001) and GRS~1915+105~(Dhawan
\etal~2000) are both resolved into elongated radio sources on
milliarcsec scales -- that is tens of A.U. -- implying collimation
angles smaller than a few degrees. Even though no collimated radio jet
has been resolved in any BHB emitting X-rays below a few per cent of
the Eddington limit, it is widely accepted, by analogy with the two
above-mentioned systems, that the flat radio spectra associated with
unresolved radio counterparts of X-ray binaries originate in
conical outflows. Yet, it remains to be proven whether such an outflow
would maintain highly collimated at very low luminosity levels, in the
so-called `quiescent' regime.

Radiatively efficient, thermal dominant X-ray states, on the contrary,
are associated with no detectable core radio emission (Fender
\etal~1999); as the radio fluxes drop by a factor up to 50 with
respect to the hard state (e.g. Corbel \& Fender 2002, Corbel
\etal~2004), this is generally interpreted as the physical suppression
of the jet taking place over this regime.

Transient ejections of optically thin radio plasmons moving away from the
binary core in opposite directions are often observed as a result of bright
radio flares associated with hard-to-thermal X-ray state transitions.

\subsection{Neutron Stars}

Low magnetic field, `atoll-type' neutron star X-ray binaries share
many X-ray spectral and timing properties with BHBs and show two
distinct X-ray states, which can be directly compared to the hard and
thermal state of BHBs; the reader is referred to Migliari \& Fender
(2006) for a comprehensive work on correlated radio and X-ray
properties of neutron stars (NSs). The main differences/similarities
can be summarized as follows: as for the BHBs, below a few per cent of
their Eddington luminosity, NSs power steady, self-absorbed
jets. Transient optically thin plasmons are ejected at higher X-ray
luminosities, in response to rapid X-ray state changes. However, {\em
NSs are less `radio loud' than BHBs.} At a given $L_{\rm X} / L_{\rm
Edd}$, the difference in radio luminosity is typically a factor
$\sim$30, which {can not be accounted for by the mass ratio alone}.
In contrast to BHs, atolls have been detected in the radio band while
in the soft X-ray state, suggesting that the jet quenching in
disk-dominated states may not be so extreme.  Like in the BHBs, the
radio luminosity of hard state NSs scales non-linearly with the X-ray
luminosity, even though the slope of the radio-to-X-ray correlation
--albeit the low statistics-- appears to be steeper for the NSs
($L_{\rm radio}\propto L_{\rm X}^b$, with $b\simeq 0.7$ for the BHs,
and $b\simeq 1.4$ for the NSs).  High magnetic field, `Z-type' NSs,
being persistently close to the Eddington accretion rate, behave
similar to the rapidly varying BHB GRS 1915+105.

Overall, these results indicate that, at the zeroth order, the process
of continuous jet formation works similarly in accreting BHs and (at
least certain) NSs. Even though quantitative differences remain to be
explained, such as the relative dimness of neutron star jets with
respect to BH jets, this indicates that the presence of an accretion
disk coupled to an intense gravitational field could be the necessary
and sufficient ingredients for steady relativistic jets to be formed,
regardless of the presence/absence of an event horizon.  The role
played by the NS dipole magnetic field in producing/inhibiting the jet
production mechanism is far from being clear, although it is generally
assumed that the higher the magnetic field, the lower the jet power.
Further radio observations of X-ray pulsars and ms accreting X-ray
pulsars are definitely needed in order to place stronger observational
constraints.  In addition, it is worth remembering that the most
relativistic Galactic jet source observed so far is a NS (Fender
\etal~2004), contrary to what it might be expected according to the
so-called `escape-velocity paradigm'.

\section{Black Holes Unified Model: Journey through an Outburst}

As proven by the case of GRS1915+105, the same source can produce
either kind of jets (steady and transient), dependent (somehow) on
the mass accretion rate, or, in other words, on the X-ray spectral
state. Thus, a fundamental question arises: {do the steady jet and the
optically-thin-discrete ejections differ fundamentally (e.g. in their
production mechanism, power content)? Or are they different
manifestations of the same phenomenon?

Fender, Belloni, \& Gallo (2004) have addressed this issue by carefully
examining the behaviour of 4 BHB systems (GRS1915+105, GX339-4,
XTEJ1859+226 and XTE J1550-564) for which nearly simultaneous
radio/X-ray coverage over different X-ray spectral states is available
in the literature. Based upon a detailed study of these 4 sources and
by comparing their properties to the whole sample of known BHBs, we
have attempted to construct a unified, semi-quantitative, model for
the disc-jet coupling in BHBs. The model can be summarized as follows
(see Figure 7 in the paper for a schematic illustrating a typical BHB
outburst cycle): Starting from the bottom right corner, the system is
in a low-luminosity hard X-ray state, producing a mildly relativistic,
persistent outflow, with a flat radio spectrum.
Its luminosity starts to increase at all wavelengths, while the X-ray
spectrum remains hard; around a few per cent of the Eddington X-ray
luminosity, a sudden transition is made (top horizontal branch) during
which the global properties of the accretion flow change from
radiatively inefficient to efficient (hard-to-thermal dominant state
transition), while a bright radio flare is observed, likely due to a
sudden ejection episode.  This is interpreted as the result of the
inner radius of a geometrically thin accretion disc moving inward, as
illustrated in the bottom panel: the Lorentz factor of the ejected
material, due to the deeper potential well, exceeds that 
of the hard state jet, causing an internal shock to propagate through
it, and to possibly disrupt it. Once the transition to the thermal
dominant state is made, no core radio emission is observed, while
large scale rapidly fading radio plasmons are often seen moving in
opposite direction with highly relativistic speed.

The bright radio flare associated with the transition could coincide
with the very moment in which the hot corona of thermal electrons,
responsible for the X-ray power law in the spectra of hard state BHBs,
is accelerated and ultimately evacuated.  This idea of a sudden
evacuation of inner disc material is not entirely new, and in fact
dates back to extensive RXTE/PCA observations of the rapidly varying
GRS~1915+105: despite their complexity, the source spectral changes
could be accounted for by the rapid removal of the inner region of an
optically thick accretion disc, followed by a slower replenishment,
with the time-scale for each event set by the extent of the missing
part of the disc (Belloni \etal~1997a,b). Subsequently,
multi-wavelength (radio, infrared, and X-ray) monitoring of the same
source suggested a connection between the rapid disappearance and
follow-up replenishment of the inner disc seen in the X-rays, with the
infrared flare starting during the recovery from the X-ray dip, when
an X-ray spike was observed.

Yet it remains unclear what drives the transition in the radio
properties after the hard X-ray state peak is reached. Specifically,
radio observations of GX~339--4, XTE J1550-564, and GRS~1915+105
indicate that in this phase the jet spectral index seems to
`oscillate' in an odd fashion, from flat to inverted to optically
thin, as if the jet was experiencing some kind of instability as the
X-ray spectrum softens.  Recent simultaneous RXTE and INTEGRAL
observations of GX~339--4 (Belloni \etal~2006) have shown that the
high energy (few 100s of keV) cutoff typical of hard state X-ray
spectra either disappears or shifts towards much higher energies
within timescales of hours ($<$8 hr) during the transition. Previous
suggestions of such behaviour were based on X-ray monitoring campaigns
with instruments such as OSSE, for which the long integration times
required in order to accumulate significant statistics did not allow
constraints to be placed on the timing and significance of rapid
changes in the X-ray spectra.  The suggestion that the so called `jet
line', where the radio flare is observed and the core radio emission
is suddenly quenched, might correspond to a peculiar region in the time
domain seems to be at odds with recent radio observations (Fender
\etal, in prep.).

Finally, there are at least a couple of recent results that might
challenge some of the premises the unified scheme is based on.  The
first one is the notion that, for the internal shock scenario to be at
work and give rise to the bright radio flare at the state transition,
whatever is ejected must have a higher velocity with respect to the
pre-existing hard state steady jet. From an observational point of
view, this was supported, on one side, by the lower limits on the
transient jets' Lorentz factors, typically higher than 2 (Fender
2003), and, on the other hand, by the relatively small scatter about the
radio/X-ray correlation in hard state BHBs~(Gallo, Fender, \& Pooley
2003). The latter has been challenged on theoretical grounds~(Heinz \&
Merloni 2004); while a recent work (Miller-Jones \etal~2006)has
demonstrated that, from an observational point of view, the average
Lorentz factors do not differ substantially between hard and transient
jets (albeit the estimated Lorentz factors rely on the assumption of
no lateral confinement).

In addition, recent high statistics X-ray observations of a hard state
BHB undergoing outburst (Miller \etal~2006) suggest that a cool, thin
accretion disc extends to near the innermost stable circular orbit
(ISCO) already during the bright phases of the hard state, that is,
prior to the horizontal branch in the top panel of Figure 7 in~Fender,
Belloni, \& Gallo.  This would challenge the hypothesis of a sudden
deepening of the inner disc potential well as the cause of a high
Lorentz factor ejection.  Possibly, whether the inner disc radius
moves close to the hole prior or during the softening of the X-ray
spectrum does not play such a crucial role in terms of jet properties;
if so, then the attention should be diverted to a different component,
such as the presence/absence, or the size (Homan \etal~2001), of a
Comptonizing corona (which could in fact coincide with the very jet
base~(Markoff, Nowak, \& Wilms 2005).  In a recent paper, Liu
\etal~(2007) give theoretical support to the observational results by
Miller \etal: within the framework of the disc evaporation model (see
references within Liu \etal~2007), it is found that a weak,
condensation-fed {\em inner} disc can be present in the hard state of
black hole transient systems for Eddington-scaled luminosities as low
as $10^{-3}$ (depending on the magnitude of the viscosity parameter).

One of the most interesting aspects of this proposed scheme --
assuming that is correct in its general principles -- is obviously its
possible application to super-massive BHs in Active Galactic Nuclei
(AGN), and the possibility of mirroring different X-ray binary states
in different classes of AGN: radio loud vs. radio quiet, LLAGN, FRI,
FRII etc. The interested reader is referred to~K\"{o}rding
\etal~(2006).

\section{Luminosity Correlations}

\subsection{Radio/X-ray}

In an attempt to assess the relation between accretion and jet
production in hard state BHBs, Corbel \etal~(2003) and Gallo, Fender
\&, Pooley (2003) have established the existence of a tight
correlation between the X-ray and the radio luminosity ($L_X$ and
$L_{R}$), of the form $L_{ R}\propto L_{X}^{0.7\pm 0.1}$. In those
works, 10 hard state systems with nearly simultaneous radio/X-ray
observations were considered; the correlation was found to hold over
more than 3 orders of magnitude in $L_{X}$, up to a few per cent of
$L_{\rm Edd}$, above which the sources enter the thermal dominant
state, and the core radio emission drops below detectable levels.
Probably the most notable implication of the non-linear scaling was
the predicted existence of a critical X-ray luminosity below which a
significant fraction of the liberated accretion power is channelled
into a radiatively inefficient outflow, rather than being dissipated
locally by the inflow of gas and emitted in the form of X-rays (this
does not necessarily imply the X-ray spectrum is dominated by
non-thermal emission from the jet, as most of the jet power may be
stored as kinetic energy; Fender, Gallo, \& Jonker 2003).

The simultaneous VLA/\cxo observations of the $10^{-8.5}L_{\rm Edd}$
BHB A0620--00 allowed us to probe the radio/X-ray correlation for BHBs
down to the truly quiescent regime. The measured radio/X-ray fluxes
seem to confirm the existence of a non-linear scaling between the
radio and X-ray luminosity in this system; with the addition of the
A0620--00 point, {$L_{\rm R}\propto L_{\rm X}^{0.58\pm0.16}$ provides a
good fit to the data for $L_{\rm X}$ spanning between $10^{-8.5}$ and
$10^{-2} L_{\rm Edd}$. The fitted slope, albeit consistent with the
previously reported value of $0.7\pm0.1$, is admittedly affected by
the uncertainties in the distance to GX339--4, for which the
correlation extends over 3 orders of magnitude in $L_{\rm X}$ and
holds over different epochs.

Even though the A0620--00 data appear to follow the hard state
radio/X-ray correlation, since 2003, when the compilation of
quasi-simultaneous radio/X-ray observations of hard state BHBs was
presented, many outliers have been found. To mention a few: XTE
J1720--318 (Chaty 2006, Brocksopp \etal~2006), SWIFT J1753.5--0127
(Cadolle-Bel \etal~2006), IGR J17497--2821 (Rodriguez \etal~2007) and
XTE J1650--500 (Corbel \etal~2004), while in the hard state, all
appear to lie significantly below the best-fitting correlation.
Beside casting doubts on the possibility of relying on the
best-fitting relation for estimating other quantities, such as
distance or mass, this suggests that the correlation between radio and
X-ray properties of outbursting BHBs is far more complex than
previously thought.

\subsection{Optical-IR/X-ray}
\begin{figure}
\centerline{\epsfig{figure=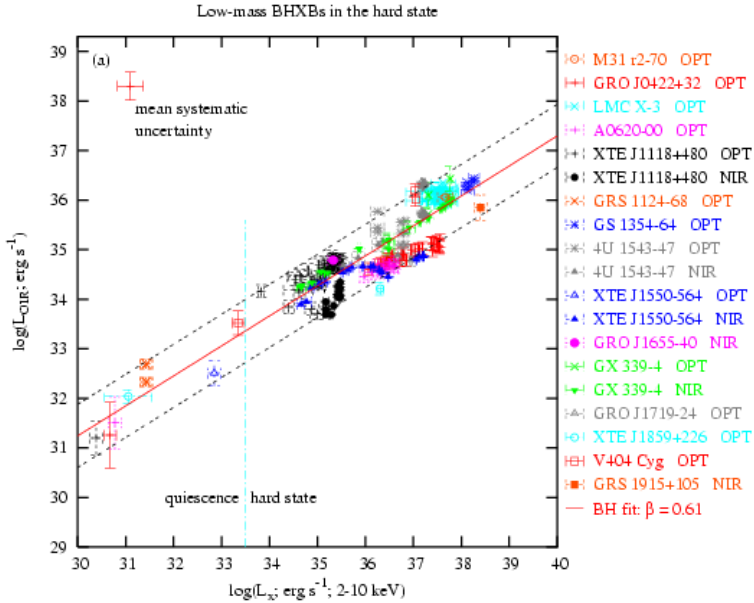,width=0.6\textwidth,angle=0}}
\caption{Optical-IR vs. X-ray correlation in hard state black
holes. From Russell \etal~(2006)}.
\label{fig:opt}
\end{figure}

In order to quantify the relative importance of jet vs. disc emission
as a function of the state and luminosity in a frequency window that
is undoubtedly crucial and yet shaped by a number of competing
mechanisms, Russell \etal~(2006) put together nearly-simultaneous
optical-IR (termed OIR) and X-ray observations of X-ray binaries, BHs,
and neutron stars.  A global correlation is found between OIR and
X-ray luminosity for low-mass BHBs in the hard state, of the form
$L_{\rm OIR}\propto L_X^{0.6}$ (see Figure~1).  This correlation holds
over eight orders of magnitude in $L_X$ and includes data from BHBs in
quiescence. A similar correlation is found in low-mass neutron star
X-ray binaries in the hard state. Similarly to what happens for the
radio emission, for thermal dominant state BHBs all of the near-IR and
some optical emissions are suppressed, indicating that the jet is
quenched during the hard-to-thermal transition. By comparing these
empirical correlations with existing models, Russell \etal~come to the
conclusion that, for the BHs, X-ray reprocessing in the disc and
emission from the jets both contribute to the optical-IR while in the
hard state, with the jet accounting for up to 90 per cent of the
near-IR emission.  In addition, it is shown that the optically thick
jet spectrum of BHBs is likely to extend to near the K band.

X-ray reprocessing dominates in the hard state neutron stars, with
possible contributions from the jets and the
viscously heated disc, only at high luminosities.

Indications for a jet contribution in the IR band also come from variability
(e.g. Homan \etal~2005, Hynes \etal~2003, 2006) and polarization studies
(e.g. Dubus \& Chaty 2006; Shahbaz \etal~2008).

\section{Spitzer Space Telescope Observations}

\subsection{Quiescent Black Holes}
The infrared spectra of BHBs with a low mass donor star are likely
shaped by a number of competing emission mechanisms, among which are:
reprocessing of accretion-powered X-ray and ultraviolet photons,
either by the donor star surface or by the outer accretion disk,
direct thermal emission from the outer disk, non-thermal synchrotron
emission from a relativistic outflow, and thermal emission from
circumbinary dust.

The Spitzer Space Telescope offers the opportunity for the first time
to identify and characterize the properties of highly sub-Eddington
Galactic BHBs in the mid-infrared band, a frequency window that is
still largely unexplored for these systems, and which can prove to be
crucial for our understanding of the overall structure of the
accretion flow in quiescence.
 While there is general agreement that in this regime the X-rays are
produced by a population of high-energy electrons near to the black
hole, the controversy comes about in modeling the contribution from
inflowing vs. outflowing particles, and their relative energy budget
(e.g. McClintock \& Remillard 2006).  There is mounting evidence that
synchrotron emitting outflows survive at very low Eddington ratios
($10^{-8.5}$\ledd; Gallo \etal~2006), and that their mechanical power
is comparable to the bolometric X-ray luminosity in higher luminosity
sources ($10^{-2}$\ledd; Gallo \etal~2005; Russell
\etal~2007). However, in the absence of information from the mid-IR
band, disentangling the relative contributions of inflow vs. outflow
to the radiation spectrum and global accretion energy budget can be
quite challenging. Estimates of the jet power based on its radiation
spectrum depend crucially on the frequency at which the flat,
partially self-absorbed radio spectrum turns and becomes optically
thin (the jet `break').
%

%
%
\begin{figure}
\centerline{\psfig{figure=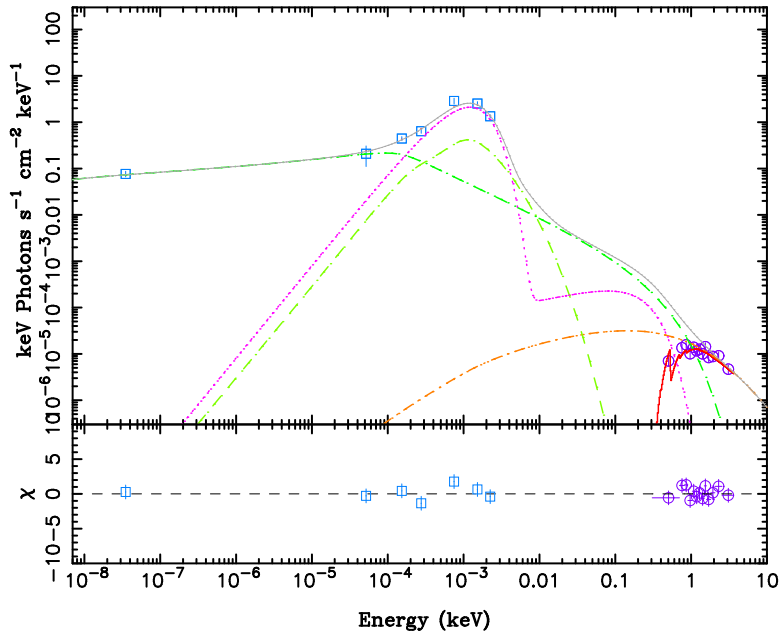,width=0.62\textwidth,angle=0}}
\caption{Broadband (VLA+Spitzer+SMARTS+Chandra) modeling of the black
hole A0620-00 ($L_X$/\ledd$\sim 10^{-8}$) with a `jet-dominated'
model. The fitted parameters land in a range of values that is
reminiscent of the Galactic Center black hole Sgr A$^{\star}$. Most
notably, the inferred ratio of the jet acceleration rate to local
cooling rates is two orders of magnitude weaker with respect to higher
luminosity, hard state sources.  From Gallo \etal~(2007).}\label{fig:sera}
\end{figure}

Because of the low flux levels expected from the jets in this regime
(10--100s of $\mu$Jy, based on extrapolation from the radio band),
combined with the companion star/outer disk contamination at near-IR
frequencies, the sensitivity and leverage offered by Spitzer IRAC and
MIPS are crucial for determining the location of the jet break.
In turn, this can constrain the possible contribution of the jet to
hard X-rays, and hence the accretion flow model.  Recent Spitzer
observations of three quiescent black hole X-ray binaries (Muno \&
Mauerhan 2006) have shown evidence for excess emission with respect to
the Rayleigh-Jeans tail of the companion star between 8--24 $\mu$m
(the disk contribution is negligible in all three cases). This excess,
which has been interpreted as due to thermal emission from cool
circumbinary material by Muno \& Mauerhan (2006), is also consistent
with the extrapolation of the measured radio flux assuming a slightly
inverted spectrum, typical of partially self-absorbed synchrotron
emission from a conical jet (Gallo \etal~2007).  If so, then the jet
synchrotron luminosity exceeds the measured X-ray luminosity by a
factor of a few in these systems. Accordingly, the mechanical power
stored in the jet exceeds the bolometric X-ray luminosity at least by
4 orders of magnitude (based on kinetic luminosity function of
Galactic X-ray binary jets;~Heinz \& Grimm 2005).

Within this framework, the broadband SED (VLA/Spitzer/SMARTS/Chandra)
of the $10^{-8.5}$\ledd\ black hole A0620--00 can be fit with a
`maximally jet-dominated' model in which the radio through the soft
X-rays are dominated by synchrotron emission, while the hard X-rays
are dominated by inverse Compton at the jet base (see Figure~2), as
described below.

The model is most sensitive to the fitted parameter $N_{\rm j}$, which
acts as a normalization, though it is not strictly equivalent to the
total power in the jets. It dictates the power initially divided
between the particles and magnetic field at the base of the jet, and
is expressed in terms of a fraction of $L_{\rm Edd}$.  Once $N_{\rm
j}$ is specified and conservation is assumed, the macroscopic physical
parameters along the jet are determined assuming that the jet power is
roughly shared between the internal and external pressures.  The
radiating particles enter the base of the jet where the bulk
velocities are lowest, with a quasi-thermal distribution. Starting at
location $z_{\rm acc}$ in the jets, a free parameter, a fraction of
85$\%$ of the particles are accelerated into a power law with index
$p$, also a fitted parameter.  The maximum energy of the accelerated
leptons is calculated by setting the acceleration rate to the local
cooling rates from synchrotron and inverse Compton radiation at
$z_{\rm acc}$.  If the acceleration process is diffusive Fermi
acceleration, the acceleration rate depends on the factor
$f=\frac{(u_{\rm acc}/c)^2}{f_{sc}}$, where $u_{\rm acc}$ is the shock
speed relative to the bulk plasma flow, and $f_{\rm sc}$ is the ratio
of the scattering mean free path to the gyro-radius.  Because neither
plasma parameter is known, the model fits for the parameter $f$, which
thus reflects the {\it efficiency of acceleration}.
In the case of A0620--00, $f$ is found to be around two orders of
magnitude lower than in higher luminosity sources. 
This `weak acceleration' scenario is reminiscent of the Galactic
Center super-massive BH Sgr A*. Within this framework, the SED of Sgr
A* does not require a power law of optically thin synchrotron emission
after the break from its flat/inverted radio spectrum~(Falcke \&
Markoff 2000). Therefore, if the radiating particles have a power law
distribution, it must be so steep as to be indistinguishable from a
Maxwellian in the optically thin regime. Something similar, albeit
less extreme, is occurring in the quiescent BHB A0620--00; either
scenario implies that acceleration in the jets is inefficient at
$10^{-9}-10^{-8}L_{\rm Edd}$.

\subsection{Ultra-compact Neutron Star X-ray Binaries}

This multi-wavelength approach has been extended to neutron star X-ray
binaries.  By means of coordinated VLA, Spitzer (IRAC), YALO SMARTS
and RXTE observations, Migliari \etal~(2006) have provided the first
spectroscopic evidence for the presence of a jet in a low-luminosity
ultra-compact {neutron star} X-ray binary (4U0614+091). The Spitzer
IRAC data (Figure 3, circles) show a neat optically {\it thin}
synchrotron spectrum ($F_{\nu}\propto \nu^{-0.6}$). Accordingly, the
inferred upper limit on the break frequency of the spectrum is
$\nu_{\rm thin}=3.7 \times 10^{13}$ Hz, which is lower than that
observed in BHBs by at least a factor of 10. This implies that, in
this system, the optically thin jet emission cannot possibly
contribute substantially to the hard X-ray emission. Assuming a
high-energy cooling cutoff at ~1 keV, the total (integrated up to
X-rays) jet power to X-ray bolometric luminosity ratio is about 5 per
cent, much lower than that inferred in BHBs.

\begin{figure}
\centerline{\psfig{figure=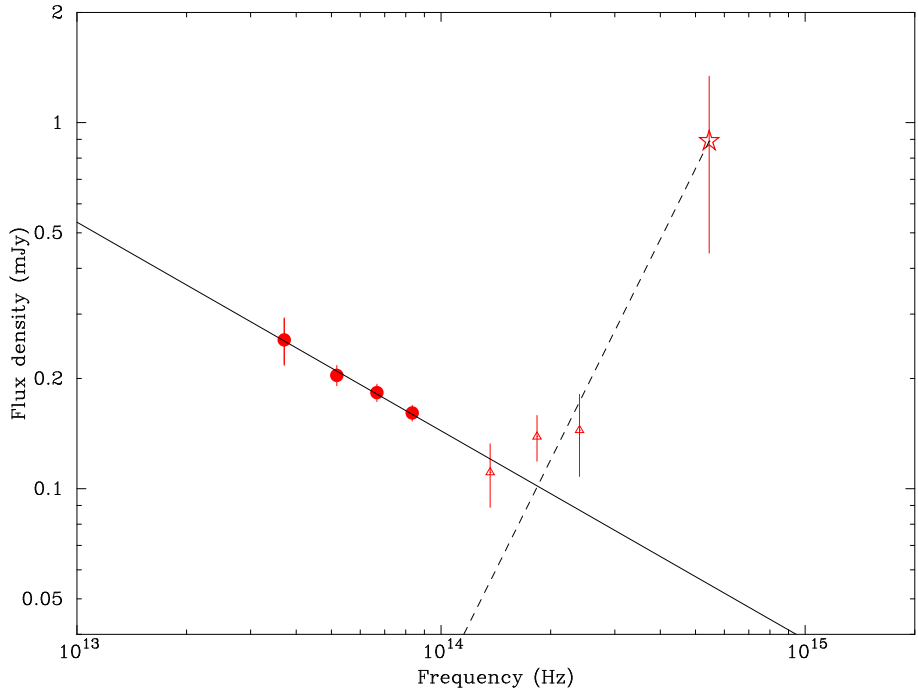,width=0.55\textwidth,angle=-90}}
\caption{Spitzer IRAC + UKIRT + SMARTS spectrum of the ultra-compact
{\it neutron star} X-ray binary 4U0614+091. The 2 components are due
to optically thin synchrotron emission from a jet (solid line) and the
tail of the disk thermal emission (dotted line). The jet is optically
thin at $\nu<4\times 10^{13}$ Hz, implying that, unlike in the black
holes, it cannot contribute significantly to the X-ray emission
(Migliari \etal~2006).}
\end{figure}

\section{Open Issues}
Rather than with a summary, I wish to conclude this review with a list of open
issues that I regard as important:
\begin{itemize}
\item
Do collimated jets survive at Eddington-scaled X-ray luminosities
lower than a few per cent?  It is often taken for granted that the
steady unresolved radio sources with flat spectra that are associated
with hard state BHBs are due to synchrotron emission from a collimated
outflow. Yet this remains to be proven from the observational side.
\item
Are there fundamental differences between BHB and NS jets?
While NSs tend to be fainter in the radio, there is no observational evidence
for a different mechanism at work in the two classes, which is to say, there
is no evidence for any mechanism that may be purely related to the direct
extraction of energy from the BH spin. 
\item
What is composition of the jets?
Sparse indications come from circular polarization studies, but conclusive
evidence is still lacking. 
\end{itemize}

\acknowledgements I wish to thank all my collaborators, and in
particular Rob Fender, Sera Markoff, James Miller-Jones, Simone
Migliari, and John Tomsick, whose work has especially contributed to
shape this review.  This work is supported by NASA through \cxo
Postdoctoral Fellowship grant number PF5-60037, issued by the \cxo
X-ray Center, which is operated by the Smithsonian Astrophysical
Observatory for NASA under contract NAS8-03060.

\end{document}